\documentclass[aps,prb,superscriptaddress,showkeys,showpacs,twocolumn,floatfix]{revtex4}

\usepackage{amssymb,amsbsy,amsmath}

\usepackage{graphicx}
\usepackage{dcolumn}
\usepackage{bm}
\usepackage{subfigure}
\usepackage{color}

\usepackage{multirow}
\usepackage{array}
\usepackage{rotating}

\newcommand{\op}[1]{%
    \fontdimen12\textfont3=2pt\fontdimen12\scriptfont3=1.4pt%
    \!\null\mathop{\vphantom{#1}\smash{#1}}\limits_{\sim}\null\!}

\newcommand{\xref}[1]{\protect\ref{#1}}
\newcommand{\figref}[1]{Fig.~\protect\ref{#1}}

\newcommand {\mofe} {\{$\textrm{Mo}_{72}\textrm{Fe}_{30}$\}}
\newcommand {\mocr} {\{$\textrm{Mo}_{72}\textrm{Cr}_{30}$\}}
\newcommand {\mov} {\{$\textrm{Mo}_{72}\textrm{V}_{30}$\}}
\newcommand {\wv} {\{$\textrm{W}_{72}\textrm{V}_{30}$\}}

\begin{document}

\title{Exchange randomness and spin dynamics in the frustrated
  magnetic Keplerate \wv}

\author{J\"urgen Schnack}
\email{jschnack@uni-bielefeld.de}
\affiliation{Fakult\"at f\"ur Physik, Universit\"at Bielefeld, Postfach 100131, D-33501 Bielefeld, Germany}
\author{Ana-Maria Todea}
\affiliation{Fakult\"at f\"ur Chemie, Universit\"at Bielefeld, Postfach 100131, D-33501 Bielefeld, Germany}
\author{Achim M{\"u}ller}
\affiliation{Fakult\"at f\"ur Chemie, Universit\"at Bielefeld, Postfach 100131, D-33501 Bielefeld, Germany}
\author{Hiroyuki Nojiri}
\affiliation{Institute for Materials Research, Tohoku University, Katahira 2-1-1, Sendai 980-8577, Japan}
\author{Steven Yeninas}
\affiliation{Ames Laboratory and Department of Physics and Astronomy, Iowa State University, Ames, Iowa 50011, USA}
\author{Yuji Furukawa}
\affiliation{Ames Laboratory and Department of Physics and Astronomy, Iowa State University, Ames, Iowa 50011, USA}
\author{Ruslan Prozorov}
\affiliation{Ames Laboratory and Department of Physics and Astronomy, Iowa State University, Ames, Iowa 50011, USA}
\author{Marshall Luban}
\affiliation{Ames Laboratory and Department of Physics and Astronomy, Iowa State University, Ames, Iowa 50011, USA}

\date{\today}

\begin{abstract}
The magnetic properties and spin dynamics of the spin frustrated
polyoxometalate {\wv}, where 30 V$^{4+}$ ions ($s = 1/2$)
occupy the sites of an icosidodecahedron, have been investigated
by low temperature magnetization, magnetic susceptibility,
proton and vanadium nuclear magnetic resonance, and theoretical
studies. The field-dependent magnetization at 0.5~K increases
monotonically up to 50~T without any sign of staircase
behavior. This low-temperature behavior cannot be explained by a
Heisenberg model based 
on a single value of the nearest-neighbor exchange coupling. We
analyze this behavior upon assuming a rather broad distribution
of nearest-neighbor exchange interactions. Slow spin dynamics of
{\wv} at low temperatures is observed from the magnetic field
and temperature dependence of nuclear spin-lattice relaxation
rate $1/T_1$ measurements. 
\end{abstract}

\pacs{75.10.Jm,75.50.Xx,75.40.Mg} \keywords{Heisenberg
model, Frustrated spin system, Numerically exact energy
spectrum, NMR}

\maketitle

\section{Introduction}
\label{sec-1}

Nanometer sized highly symmetric polyoxometalate molecules
constitute a fascinating class of molecular materials.
\cite{MPP:CR98,MSS:ACIE99,MKD:CCR01,MLS:CPC01,Mue:Sci03,KMS:CCR09,KTM:DT10,FuW:RMP13}
The series of Keplerate\footnote{The
    term Keplerate was coined to describe molecular structures
that contain Platonic and Archimedean solids 
because of the use of such structures in an early model of the
solar system by Johannes Kepler.\cite{MKB:ACIE98}}
 clusters
\mofe,\cite{MSS:ACIE99,MLS:CPC01}
\mocr,\cite{TMB:ACIE07}
\mov,\cite{MTS:AC05,BKH:CC05}
and
\wv\cite{TMB:CC09}
is from a magnetism point of view of special interest since in
these bodies paramagnetic ions occupy the vertices of a nearly
perfect icosidodecahedron -- one of the Archimedian
solids. 

\begin{figure}[ht!]
\centering
\includegraphics*[clip,width=55mm]{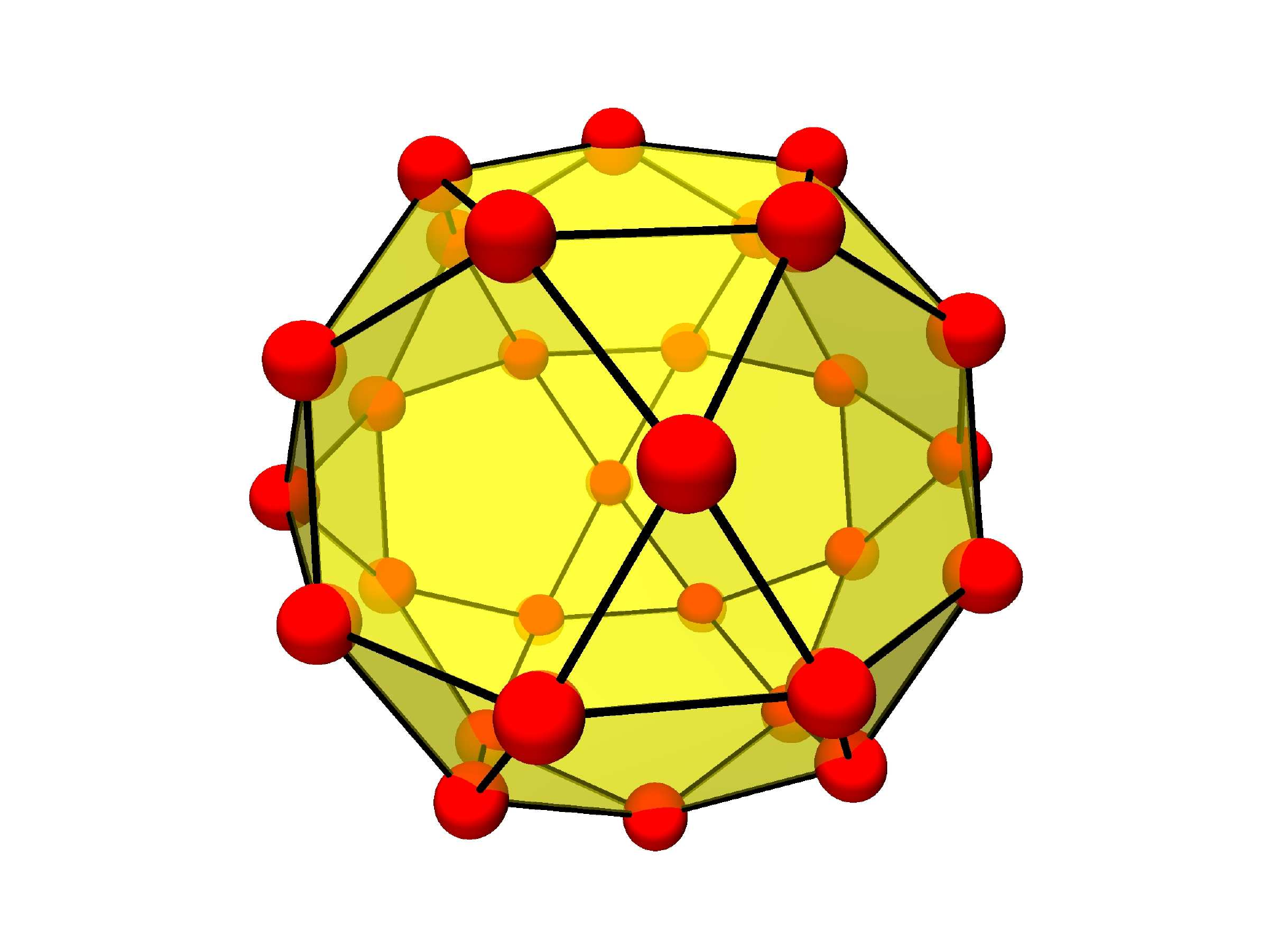}
\caption{(Color online) The core structure of the presently
  investigated Keplerate 
  cluster is an 
  icosidodecahedron. The bullets represent the 30 spin sites
  and the edges correspond to the 60 exchange interactions between
  nearest-neighbor spins.}
\label{w72v30-f-1}
\end{figure}

Figure~\xref{w72v30-f-1} shows the structure of the
icosidodecahedron: spin sites are displayed by bullets, 
edges represent interaction pathways between nearest-neighbor
spins, see the Heisenberg model Hamiltonian given in
\eqref{E-2-1}.  If such
interactions are of antiferromagnetic nature, i.e. favor
antiparallel alignment in the ground state, a magnetic structure
that consists of triangles is said to be frustrated.
\cite{Ram:ARMS94,Gre:JMC01,Sch:DT10} In this respect the
icosidodecahedron belongs to the archetypical class of
frustrated spin systems made of corner-sharing triangles as does
the two-dimensional kagom\'e lattice
antiferromagnet.\cite{SML:PRL00,Moe:CJP01,Atw:NM02,SRM:JPA06,RLM:PRB08,Moe:JPCS09} 
Compared to other antiferromagnetically coupled spin systems,
such as spin rings for instance, these structures possess
unusual features generated by the frustration: (1) many
low-lying singlet states below the lowest triplet excitation, (2)
an extended plateau of the magnetization at one-third of the
saturation magnetization when plotted versus field at low
temperatures, and (3) a large magnetization jump to saturation,
again as function of applied magnetic
field.\cite{SSR:EPJB01,SHS:PRL02,SNS:PRL05,SRM:JPA06} The last
feature is intimately connected with a huge magnetocaloric
effect.\cite{Zhi:PRB03,SSR:PRB07}  It is the hope that valuable insight
about the physics of lattices such as the kagom\'e lattice can
be gained by studying the finite-size bodies. 

Surprisingly, it turned out that the low-temperature
magnetization versus external field $B$ of \mofe\ and \mocr\
deviates substantially from 
the expectation for a regular icosidodecahedron with a single 
nearest-neighbor exchange interaction.\cite{SPK:PRB08} Although
the temperature dependence of the weak-field 
susceptibility could be rather well reproduced by a Heisenberg model
with a single exchange constant, the low-temperature
magnetization could not. Later investigations revealed 
a strong dependence at low temperatures $T$ of the differential
susceptibility $\partial {\mathcal M}/\partial B$ on $T$ and
$B$.\cite{SPK:PRB08,SFF:JPCM:10} These results were 
explained in the framework of classical spin dynamics by
assuming a distribution of random nearest-neighbor exchange
interactions.\cite{SPK:PRB08,SFF:JPCM:10} This means that the exchange
interactions between nearest neighbor spins of each molecule in
the bulk sample are selected from a random distribution whose
mean exchange 
constant reproduces the high-temperature results. The mere fact
that exchange interactions of a real substance might fluctuate
around a mean value might not be surprising. What is indeed
surprising is the large spread of values that had to be
assumed: the exchange interactions $J$ had to vary from half to
twice the mean $J$ (in the non-symmetric
distribution).\cite{SPK:PRB08} 

In this article we discuss the magnetic properties of a 
recent member of the family of Keplerates, \wv, where 30 V$^{4+}$
ions (spins $s=1/2$) occupy the sites of the icosidodecahedron.
In a previous work, the
high-temperature ($T>70$~K) part of the susceptibility data
measured at $B=0.5$~T could be successfully explained using the 
Quantum Monte Carlo method (QMC) on choosing the
antiferromagnetic nearest-neighbor exchange constant $J=-57.5$~K
and the spectroscopic splitting factor $g=1.95$.\cite{TMB:CC09} 
These numerical values are associated with a Heisenberg Hamiltonian written as 
\begin{eqnarray}
\label{E-2-1}
\op{H}
&=&
- 2 J 
\sum_{<i,j>}\;
\op{\vec{s}}_i \cdot \op{\vec{s}}_j
+
g\, \mu_B\, B\,
\sum_{i}\;
\op{s}^z_i
\ .
\end{eqnarray}
Here $<i,j>$ indicates a sum over distinct nearest-neighbor
pairs and $\mu_B$ denotes the Bohr magneton. The QMC method
could not be used to establish the magnetic properties of this
system below 70 K due to the well-known negative-sign problem
for frustrated spin systems.\cite{HeS:PRB00}  Herein lies the important
advantage of \wv: Due to the small spin quantum number, $s=1/2$,
of the individual V$^{4+}$ ions, rather than using classical
methods, highly accurate quantum calculations can be performed
despite the huge size ($2^{30} = 1,073,741,824$) of the Hilbert
space dimension for this system. As shown below, we are able to
calculate the relevant thermodynamic observables as functions of
both temperature and applied field by means of the
Finite-Temperature Lanczos Method
(FTLM).\cite{PhysRevB.49.5065,ScW:EPJB10} 
In particular, we are able to show that calculations based on
\eqref{E-2-1} on choosing a single value of $J$ do not agree
with the measured susceptibility data below
15~K and especially the 
field-dependence of the low-temperature magnetization. However,
we are able to achieve reasonable agreement between theory and
experiment upon generalizing \eqref{E-2-1} so that the numerical value
of the exchange constant for any given nearest-neighbor pair is
selected using a broad probability distribution constrained so
that the mean value equals -57.5~K. Our analysis allows us to
estimate the magnitude of the exchange disorder in the
compound. As for the other Keplerates, that magnitude is
surprisingly large, given the fact that x-ray structure
investigations point to a highly symmetric exchange network. 

The article is organized as follows. In Sec.~\xref{sec-2} we
briefly provide experimental and theoretical details. In
Sec.~\xref{sec-3} we provide our experimental results
(susceptibility versus $T$, low temperature magnetization versus
$B$, and NMR measurements) and wherever possible compare with
our model calculations.  The article closes with a brief
summary.

\section{Experimental and theoretical methods}
\label{sec-2}

Polycrystalline samples of  \wv = 
K$_{14}$(VO)$_2$[K$_{20}\subset$\{(W)W$_5$O$_{21}$(SO$_4$)\}$_{12}$(VO)$_{30}$(SO$_{4}$)(H$_{2}$O)$_{63}$]
$\cdot$150 H$_{2}$O were synthesized using the procedure given in
Ref.~\onlinecite{TMB:CC09}. The temperature dependence of the
magnetic susceptibility $\chi={\mathcal M}/B$ for fixed $B=0.1$~T was
measured at Ames Laboratory in a temperature range of 1.9-300~K
using a Quantum Design Magnetic Properties Measurement System.  
Magnetization measurements were made at the high-field
facilities at the Institute for  Materials Research (IMR) of
Tohoku University.  Using pulsed fields, values of the
magnetization were achieved for field strengths up to 50~T. Two
types of cryostats, a conventional $^4$He bath type cryostat and a
gas-flow type cryostat, were used for the low  and high
temperature ranges, respectively. Nuclear magnetic resonance
(NMR) measurements were carried out at Ames Laboratory on $^1$H
($I =1/2, \gamma/(2\pi) = 42.5775$~MHz/T) and $^{51}$V ($I = 7/2,
\gamma/(2\pi) = 11.193$~MHz/T) by using an in-house phase-coherent
spin-echo pulse spectrometer.  The NMR spectra were obtained
either by Fourier transform of the echo signal or by sweeping
$B$. The NMR echo signal was obtained by  means of a Hahn echo
sequence with a typical $\pi/2$  pulse length of 1.0~$\mu$s. The
nuclear spin-lattice relaxation time $T_1$ was measured by the
saturation method with the frequency at the highest peak
position of the NMR spectrum.

Our numerical FTLM calculations for the Heisenberg model were
performed on a supercomputer. We employed the SGI Altix 4700 
as well as the SuperMIG cluster
at the German Leibniz Supercomputing Center using openMP
parallelization with up to 510 cores.

\section{Results and interpretation}
\label{sec-3}

\subsection{Weak-field susceptibility}
\label{sec-3-1}

\begin{figure}[ht!]
\centering
\includegraphics*[clip,width=75mm]{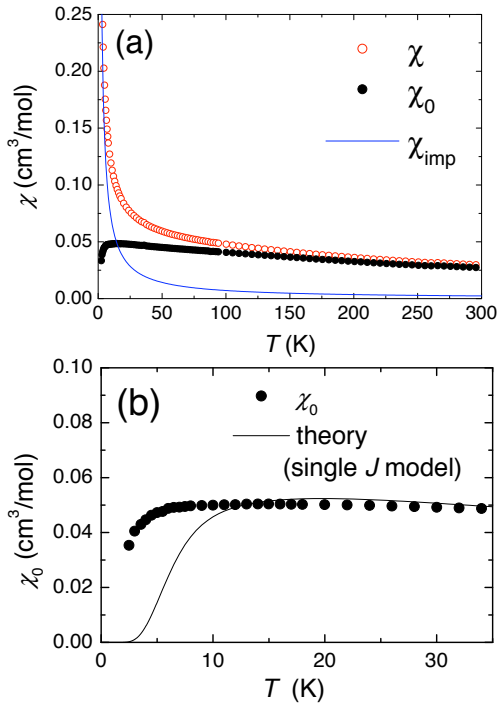}
\caption{(Color online) Molar magnetic susceptibility at $B=0.1$~T as function of
  temperature: (a) measured data of the
  crystalline compound (red circles),  
  contribution of
  two lattice VO$^{2+}$ (solid blue curve), and intrinsic susceptibility
  $\chi_{\text{0}}$ 
  of the Keplerate cluster \wv\ 
  (black circles), compare
  Ref.~\onlinecite{TMB:CC09}. (b) intrinsic susceptibility
  $\chi_{\text{0}}$ (black circles) and theoretical
  susceptibility using the single-J model (see text).} 
\label{w72v30-f-2}
\end{figure}

Figure \xref{w72v30-f-2} (a) shows the temperature dependence of
the molar susceptibility  $\chi$ measured at a field of
0.1~T. The 
measured data (open red circles, corrected for
  the effects of diamagnetism and
  temperature independent paramagnetism) increases
monotonically  with 
decreasing $T$ and obeys Curie's law for the lowest $T$. The latter
behavior is due to the  presence of the V$^{4+}$ spins of two
uncorrelated vanadyl ions, VO$^{2+}$, per formula unit located
between  the  
individual \wv\ molecules.\cite{TMB:CC09} The contribution of
these ions, denoted by $\chi_{\text{imp}}$, is
shown by the solid curve in \figref{w72v30-f-2} (a).
Subtracting that contribution from the measured susceptibility yields
values of the intrinsic susceptibility, to be denoted by
$\chi_{\text{0}}$,  of the \wv\ molecules. This data (solid black
circles) shows a broad peak around 20~K and a rapid decrease
below approximately 10~K, indicating a singlet ground state for
the \wv\ molecule. The data for $\chi_{\text{0}}$ is 
shown in an expanded scale in \figref{w72v30-f-2} (b). The
solid curve corresponds to the results for $\chi_{\text{0}}$ as
obtained using the model Hamiltonian of \eqref{E-2-1} for the
above values $J =-57.5$~K and $g = 1.95$ (the ``single-J
model"). Good agreement  between theory and experiment is
obtained only for $T > 15$~K. Below that temperature the two
data sets depart markedly from each other.

\subsection{Low temperature magnetization}
\label{sec-3-2}

\begin{figure}[ht!]
\centering
\includegraphics*[clip,width=75mm]{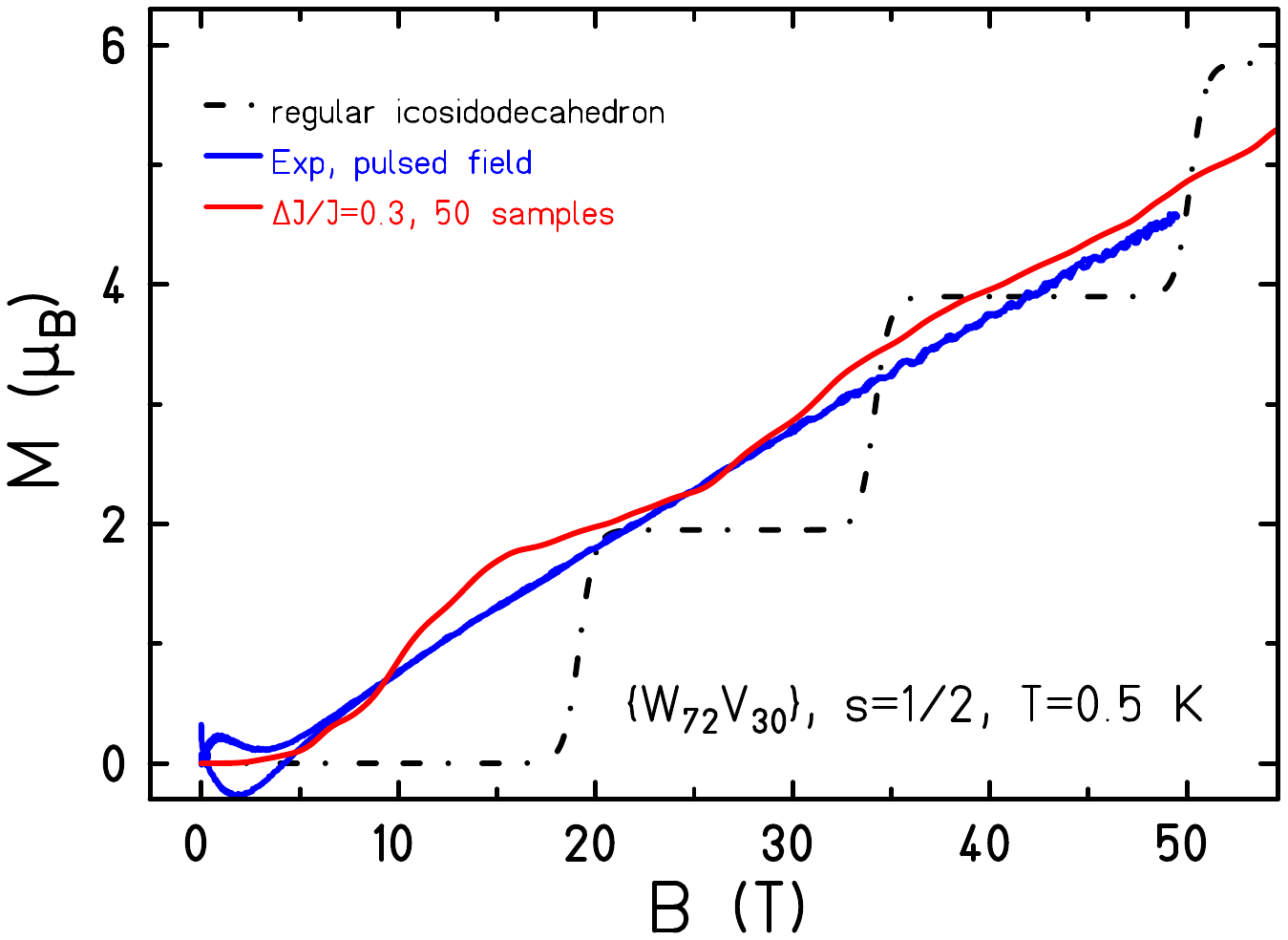}
\caption{(Color online) Intrinsic magnetization
  of the Keplerate anion \wv\ as function of applied
  field for $T=0.5$~K. Pulsed-field data are given by the blue
  curve, 
  the theoretical magnetization for the single-J model is
  shown by the black dashed-dotted curve, and for the multiple-J
  model with 
  $\Delta J/\overline{J}=0.3$ by the
  red curve.} 
\label{w72v30-f-3}
\end{figure}

In \figref{w72v30-f-3} the solid blue curve corresponds to our data
for the intrinsic magnetization versus external
  field as obtained by 
pulsed-field measurements at 0.5~K. The experimental data is corrected
  for two lattice VO$^{2+}$. The black dashed-dotted
curve is the 
result obtained for the  single-J model. Note the striking
staircase behavior of the theoretical curve for this
temperature. Surprisingly, the experimental data shows no signs
of staircase behavior.  This negative result is similar to that
for \mofe\ and \mocr. The latter systems possess much smaller
exchange couplings so one could imagine that the expected steps
are more readily washed out due to structural fluctuations or
possibly as a result of single-ion anisotropy or
Dzyaloshinskii-Moriya interactions.\cite{HaS:JPSJS04} However,
for the present system the magnetization steps of the theory are
so well separated that one would expect to see at least a hint
of them. Moreover, single-ion anisotropy is absent for V$^{4+}$
ions with spin $s=1/2$.

\subsection{Distribution of nearest-neighbor couplings}
\label{sec-3-3}

In view of the above striking discrepancies between experiment
and theory, and in  particular the failure of the single-J
model at low $T$ values (see also
  Ref.~\onlinecite{SPK:PRB08}) we assume that this might be due
  to possible low-temperature structural
    distortions  
 (as e.g. observed in some kagome
  lattices\cite{KNO:JPSJ04,RTT:CM11,YMT:JMC12}), the
  sensitivity of the exchange interactions on the  
local environment (highly charged anionic and
  cationic lattice with many dipolar crystal water
  molecules that possibly order). It is
very important at this point to understand that a  symmetric
structural distortion that would express itself in just a few
distinct exchange interactions would only alter the staircase in
a minor way but could not wash it out  completely.  For this to
happen one needs a very large number of different interactions
within each and every molecule. 

\begin{figure}[ht!]
\centering
\includegraphics*[clip,width=75mm]{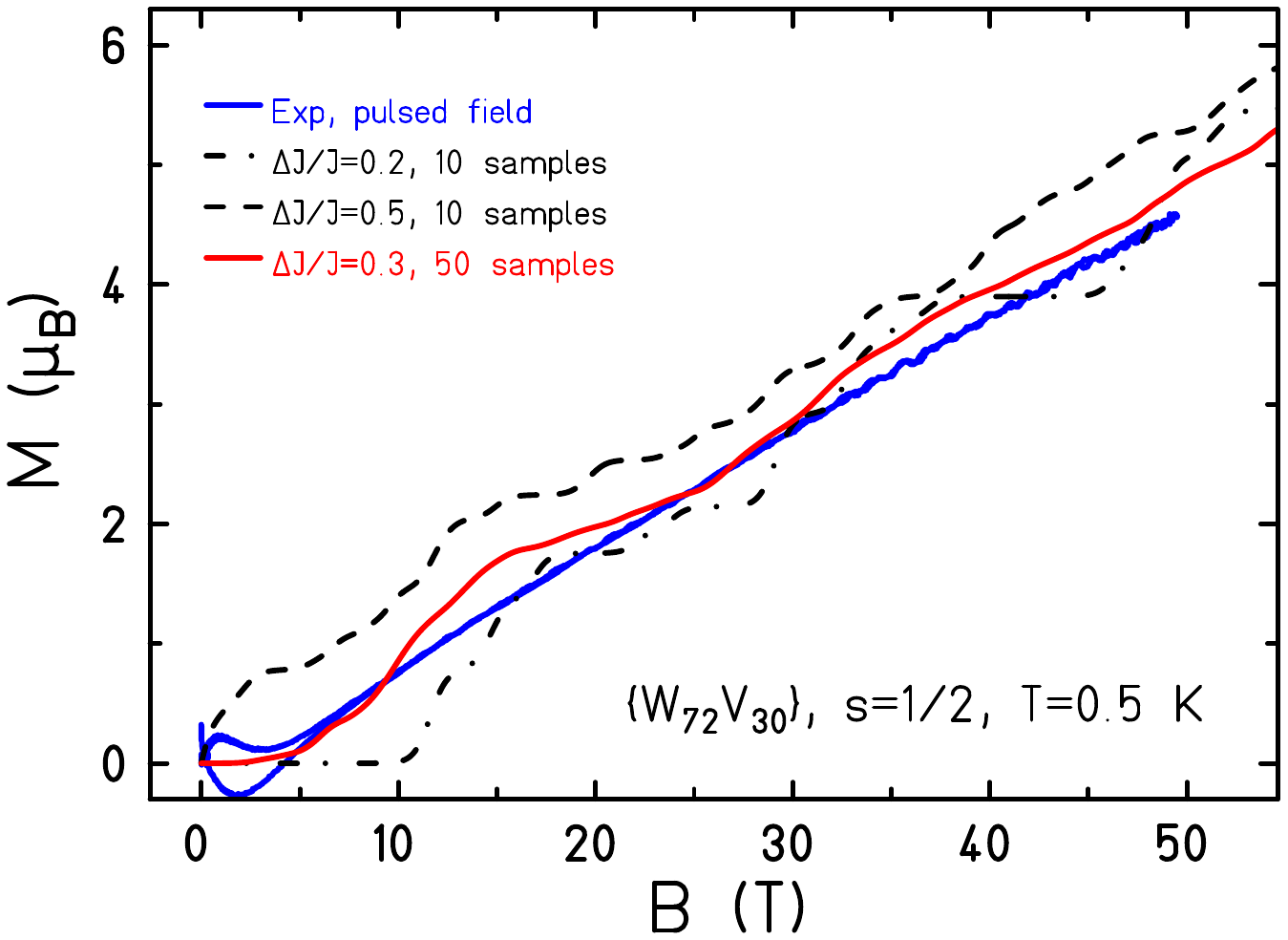}
\caption{(Color online) Intrinsic magnetization
  of the Keplerate anion \wv\ as function of applied field
  for $T=0.5$~K. Curves for
  various exchange variations $\Delta J/\overline{J}$ are
  compared to the pulsed field data.} 
\label{w72v30-f-4}
\end{figure}

Since the calculations are very computer-intensive we aim for a
coarse estimate of  the size of the exchange variation.  To this
end we used a flat distribution with  $\overline{J}-\Delta J < J
< \overline{J}+\Delta J$
and evaluated the magnetic observables for $\Delta
J/\overline{J}=0.1, 0.2, 0.3, 0.5$, with the mean $\overline{J}=-57.5$~K.
Figure~\xref{w72v30-f-4} shows the magnetization versus field for 
various choices of $\Delta J$. As can be deduced already from a
small number of samples, step-like behavior persists for $\Delta
J/\overline{J}=0.2$ and below. It turns out that the data for
$\Delta J/\overline{J}=0.3$ comes closest to the experimental
data, and that choice is shown as the red curve in
\figref{w72v30-f-3}. Since we averaged over only 50 samples that
curve is still somewhat wiggly but it is sufficiently converged
to warrant our conclusions. 

\begin{figure}[ht!]
\centering
\includegraphics*[clip,width=75mm]{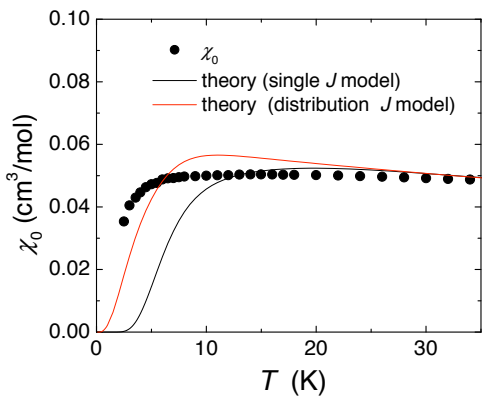}
\caption{(Color online) Molar magnetic susceptibility, $\chi_0$,
  at $B=0.1$~T as function of temperature. The curves are the result of our
  simulations for $\Delta J/\overline{J}=0.3$ and 0.}
\label{w72v30-f-5}
\end{figure}

Finally, shown in \figref{w72v30-f-5} are the results for the
intrinsic susceptibility $\chi_{\text{0}}$  as  obtained from
our measurements, 
for the single-J model, as well as for exchange variation with
$\Delta J/\overline{J}=0.3$.  
We conclude that the introduction of exchange variation yields
results that are in reasonably good agreement with our
experimental susceptibility data.

\subsection{NMR measurements}
\label{sec-3-4}

\subsubsection{$^1$H-NMR spectrum}

\begin{figure}[ht!]
\centering
\includegraphics*[clip,width=75mm]{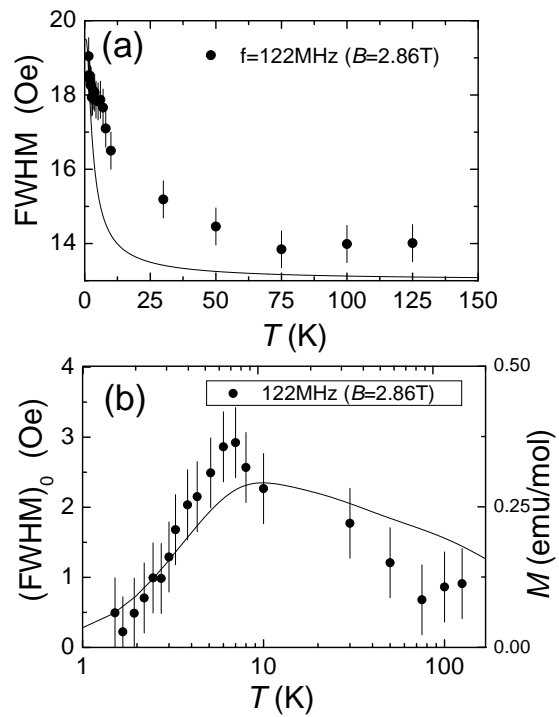}
\caption{(a) $^1$H-NMR line width (FWHM) at $B = 2.86$~T as a
  function of temperature. The solid curve shows the fitting
  result $a + b{\mathcal M}_{\text{imp}}$ (see text).  (b) $T$-dependence of
  the intrinsic line width, (FWHM)$_0$, given by $c{\mathcal
  M}_0$.  The solid line is the 
  calculated $T$-dependence of the magnetization with the
  exchange disorder $\Delta J/\overline{J}=0.3$ for the same
  field.} 
\label{w72v30-f-6}
\end{figure}

$^1$H-NMR spectra were measured for a magnetic field $B =
2.86$~T as  a function of temperature from 1.8 to
150~K. A single NMR line is observed and the line  broadens with
decreasing temperature as shown in \figref{w72v30-f-6} (a) where
the temperature dependence of the line width (full width at half
maximum, FWHM) is plotted. The FWHM in \wv\ can be expressed
as the sum 
$a + b{\mathcal M}_{\text{imp}} + c{\mathcal M}_0$. The constant
term $a$ originates from  nuclear-nuclear dipolar interactions of
the order of 10~Oe, and the second and third terms  
represent the dipolar field contributions produced by the V$^{4+}$
spins of the lattice VO$^{2+}$
ions and the intrinsic magnetic molecule, respectively. The
quantities ${\mathcal M}_{\text{imp}}$ and ${\mathcal M}_0$ are
the  corresponding magnetizations and $b$ and $c$ are parameters
related to the average dipolar   hyperfine coupling associated
with the two sets of V$^{4+}$ spins. In particular ${\mathcal
  M}_{\text{imp}}$ is   proportional to the standard expression 
$\text{tanh}(\mu_B B/k_B T)$ for independent spins $s =
1/2$. The  increase of the FWHM at low temperatures is well
reproduced by  
the above expression on choosing $a\sim 13$~Oe and $b\sim
6$~Oe/$\mu_B$ as shown by the solid  lines in
\figref{w72v30-f-6} (a). By subtracting these contributions from
the total FWHM, we obtain the  
intrinsic line width, to be denoted by (FWHM)$_0$, that is
proportional to ${\mathcal M}_0$. That data is  
shown in \figref{w72v30-f-6} (b) and it has a broad peak around
10~K. The solid curve in \figref{w72v30-f-6} (b)
corresponds to the theoretical  
result for ${\mathcal M}_0$ for exchange disorder $\Delta
J/\overline{J}=0.3$ and for an external magnetic field of
$B=2.86$~T. The experimental data is in reasonable agreement 
with the theoretical result. In this context one
  should also keep in mind that the NMR relaxation deviates from
  a single exponential behavior due to many inequivalent proton
  positions of this water rich substance, see also
  Refs.~\onlinecite{JPV:JAP02,LMC:PRB07}.

\subsubsection{$^{51}$V-NMR spectrum}

\begin{figure}[ht!]
\centering
\includegraphics*[clip,width=75mm]{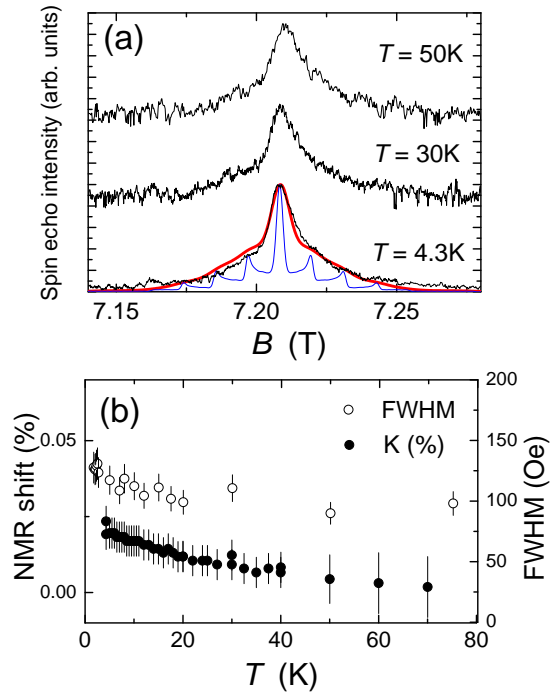}
\caption{(Color online) (a) Typical $^{51}$V-NMR spectra measured at $f
  =80.7$~MHz for various temperatures.  The blue curve shows a
  typical powder-pattern NMR spectrum with $\nu_Q =
  0.25$~MHz. The red curve is the simulated NMR spectrum with
  $\sim 40$~\% distribution of $\nu_Q$. (b) Temperature
  dependence of $^{51}$V NMR shift and line width (FWHM).} 
\label{w72v30-f-7}
\end{figure}

Figure \xref{w72v30-f-7} (a) shows typical $^{51}$V-NMR spectra
measured at $f =80.7$~MHz for various temperatures. The
$^{51}$V nucleus has nuclear spin $I = 7/2$ so that one expects
seven  quadrupole-split lines. These spectra can be calculated
using a simple nuclear spin  Hamiltonian \cite{CBK:1977}
\begin{eqnarray}
\label{E-3-1}
\op{H}
&=&
\gamma \hbar \vec{I}\cdot\vec{B}_{\text{eff}}
+
\frac{h \nu_Q}{6}
\left(
3 I_z^2-I(I+1)
\right)
\ ,
\end{eqnarray}
where $\vec{B}_{\text{eff}}$ is the effective field 
(the sum of the external field $\vec{B}$ and the hyperfine field
$\vec{B}_{\text{hf}}$) at the V$^{4+}$ site, $h$ is Planck's
constant, and $\nu_Q$ is the nuclear quadrupole frequency. The latter
quantity is proportional  
to the Electric Field Gradient (EFG) at the V$^{4+}$ site (an
asymmetric parameter of EFG is  
assumed to be zero for simplicity). The blue curve in
\figref{w72v30-f-7} (a) 
shows a typical powder-pattern spectrum calculated from the
simple Hamiltonian with $\nu_Q = 0.25$~MHz. In order to  
reproduce the observed spectrum, one needs to introduce a
distribution of $\nu_Q$. By taking  $\sim 40$~\% distributions 
($\Delta \nu_Q \sim 0.1$~MHz), one can well reproduce the observed spectrum as 
shown by the red curve in the figure. The wide distribution of
$\nu_Q$, which reflects the  
distribution of the EFG, indicates a high degree of
inhomogeneity of the local environments of the vanadium spins,
which is mainly due to the large
  difference between the distances of the two axial V - O bonds,
  one longer single and one shorter terminal double bond 
(see crystal structure in Ref.~\onlinecite{TMB:CC09}).

The temperature dependence of the NMR shift (denoted by $K$) and
FWHM for the  
$^{51}$V NMR spectrum is shown in \figref{w72v30-f-7}~(b); both
show a very weak temperature dependence. 
The hyperfine coupling constant can be estimated from the $T$-dependence of 
$K$ by comparing with the $T$-dependence of $\chi_0$. The
resulting value is very small, less than 
100~Oe/$\mu_B$. Usually the hyperfine coupling  
constant for V$^{4+}$ ions will be dominated by
core-polarization, of the order of
$-100$~kOe/$\mu_B$,\cite{FNK:PRB07}  which is three orders of
magnitude larger than our result. 
One might attempt to explain the very small value of $K$ by
suggesting that the separate contributions to
the total hyperfine field from, first, intra-atomic interactions of
$d$ electron orbitals and dipolar hyperfine field and, second,   
the transferred hyperfine field due to other V ions, nearly
cancel.
We believe, that this is not the case because $1/T_1$ of
$^{51}$V is almost 50 times smaller than that of $^1$H as will
be shown in the following section. Since $1/T_1$ is given by a
sum of all contributions of the hyperfine fields and each
contribution is proportional to the square of the hyperfine
coupling constant, it follows that $1/T_1$ of
$^{51}$V should be larger than that of $^1$H even if the total
hyperfine field is small due to the cancellation. At present we
do not have a clear explanation for the small hyperfine field on
V in \wv. It should however be noted that 
the observed V NMR signal is not coming from non-magnetic V impurities as can be 
clearly seen in the $1/T_1$ data where
$^{51}$V-$T_1$ shows a similar $T$-dependence  with that of
$^1$H-$T_1$.

\subsubsection{Nuclear spin lattice relaxation rate $1/T_1$}

To investigate the dynamical properties of the V$^{4+}$ spins, we
have carried out $^1$H-$T_1$  measurements in the temperature
range $1.5 - 100$~K. We find that $1/T_1$ is almost  
independent of temperature above $\sim 30$~K. Below $\sim 30$~K,
with decreasing temperature, $1/T_1$ starts to increase and then
shows a peak around 6~K at $B = 1.17$~T. As the external
magnetic field increases, the peak temperature of $1/T_1$ shifts
to higher temperatures and at the same time the peak height
decreases.  

In recent years it has been found,\cite{BLL:PRB04,SCL:PRL05,BFL:06} that
in many antiferromagnetic rings and clusters of spins $s>1/2$
the quantity  $1/T_1$ is well approximated by the formula 
\begin{eqnarray}
\label{E-3-3}
\frac{1}{T_1}
&=&
A\, \chi_0\, T\,
\frac{\Gamma}{\Gamma^2 + \omega_L^2}
\ ,
\end{eqnarray}
where the electronic correlation frequency $\Gamma$ is given by
a near-universal power law temperature dependence with an
exponent in the range $3.5\pm 0.5$, and  $A$ is a 
fitting constant independent of both $H$ and $T$ related to the
hyperfine field. The form of \eqref{E-3-3} is due to the fact
that the damping of the equilibrium fluctuations of the
$z$-component of  the total magnetization, $S_z$, is
monoexponential (Markovian). Specifically this is due to a dynamical
decoupling of $S_z$ from the slow 
degrees of freedom originating from the discreteness of the
energy spectrum and the conservation law $[S_z,H_0]$, where
$H_0$ denotes the 
Heisenberg model Hamiltonian of exchange coupled ion
spins.\cite{Rou:PRB07}  It follows from Eq. (3) that the quantity 
$1/(T_1 T \chi_0)$ has a maximum as a function of $T$ and fixed
$B$ when $\Gamma = \omega_L$, and its maximum value is
proportional to $1/B$. A quantitative microscopic theory
explaining all of these features has been given in 
Ref.~\onlinecite{RLB:PRB09} for spins $s>1/2$ and in particular the
numerical value of the power law exponent originates from
one-phonon acoustic processes.  It is significant that, although
$s = 1/2$, our data for \wv\ below 20~K exhibits the very same
behavior, and we find $A = 4.7\times 10^{11}$~rad Hz$^2$~mol/(K
cm$^3$) and
$\Gamma = 6.3\times 10^{5} T^{3.5}$~rad~Hz, where $T$ denotes
the temperature in units of Kelvins. In \wv\ we
find that the paramagnetic fluctuations are dominant for the
high-temperature range where the fluctuation frequency is
independent of $T$. 

\begin{figure}[ht!]
\centering
\includegraphics*[clip,width=75mm]{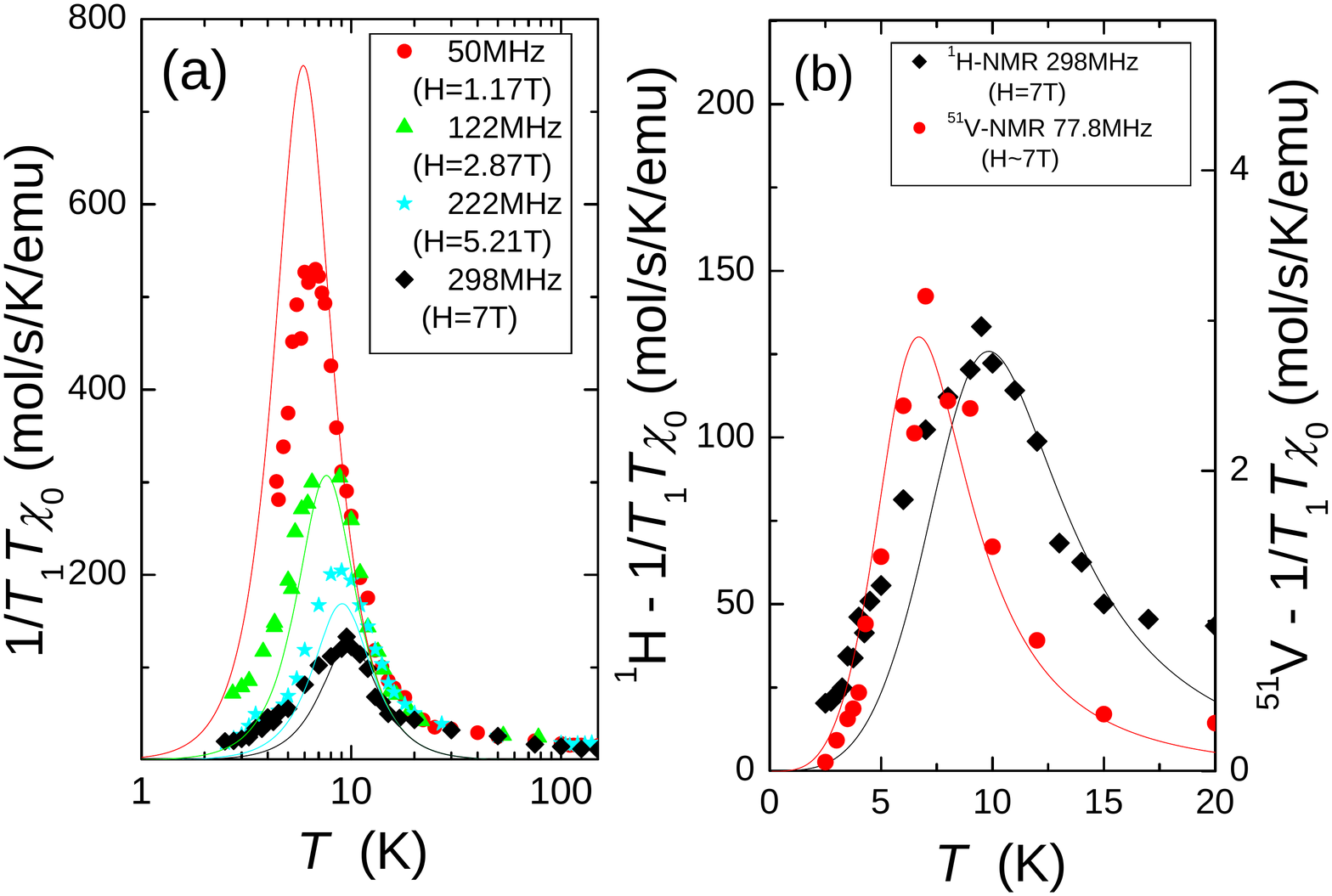}
\caption{(Color online) (a) $T$-dependence of $1/(T_1 T \chi_0)$ for
  $^1$H-NMR. Solid lines are theoretical curves calculated by
  using  \eqref{E-3-3}. (b) $T$-dependences of $1/(T_1 T
  \chi_0)$ for $^1$H-NMR ($f = 298$~MHz)  and $^{51}$V-NMR ($f =
  77.8$~MHz) at the same magnetic field $B\approx 7$~T (that is,
  same electron Larmor frequency $\omega_e$).}
\label{w72v30-f-8}
\end{figure}

In general, one expects that \eqref{E-3-3} should be
supplemented by a second Lorentzian, where the nuclear Larmor
frequency $\omega_L$ is replaced by the electron Larmor
frequency $\omega_e$. However, if $\Gamma$ is of the order of
$\omega_L$, the contribution of the second Lorentzian is
negligible. To confirm this experimentally, we performed $1/T_1$
measurements on two different nuclei, $^1$H and $^{51}$V. 
Values of $T_1$ for both nuclei were  measured at $f= 77.8$~MHz
for $^{51}$V-NMR and at 298~MHz for $^1$H-NMR, respectively, so
as  to achieve the same value of $\omega_e$, that is the same
magnetic field. The two sets of  experimental data are shown in
\figref{w72v30-f-8} (b) and they are both successfully fitted by
\eqref{E-3-3} (solid curves). If in \eqref{E-3-3} one was to
include $\omega_e$, the peak position of $1/T_1$ must be  
observed at the same temperature. These measurements provide a
firm confirmation that  
the fluctuation frequency of V$^{4+}$ spins slow down to the order
of MHz.

In antiferromagnetic rings and clusters of spins $s >1/2$ the
peak in $1/(T_1 T \chi_0)$ is usually observed for temperatures
of the order of the exchange coupling
constant.\cite{BLL:PRB04,Rou:PRB07} 
In Ref.~\onlinecite{RLB:PRB09} this is explained by the fact
that the relaxation mechanism when $s > 1/2$ is governed by the
quasi-continuum portion of the quadrupolar fluctuation spectrum
and not by the lowest excitation lines.
By contrast, in \wv\ we find that the peak temperature ($\approx
6-10$~K) is an order of magnitude smaller than
$\overline{J}=-57.5$~K.
We speculate that this difference in behavior could be explained
by invoking the modifications of the microscopic theory proposed
in Ref.~\onlinecite{RLB:PRB09} for spins $s = 1/2$, namely by
using a dipolar channel or fluctuating Dzyaloshinskii-Moriya
interactions. However, this remains to be confirmed by detailed
calculations that are outside the scope of the present work.

\section{Summary}
\label{sec-4}

We have investigated magnetic properties and spin dynamics of
\wv\ magnetic clusters by low temperature magnetization,
magnetic susceptibility, proton and vanadium NMR measurements,
and theoretical studies. Our most striking experimental finding
is that the field-dependent magnetization at 0.5~K, as obtained
using a pulsed magnetic field, increases monotonically up to
50~T without showing any sign of staircase behavior. This is 
contrary to the predictions of any model based on a single value
of the nearest-neighbor exchange coupling. Also we find that a
``single-J model" fails to describe the temperature
dependence of the intrinsic weak-field magnetic susceptibility
$\chi_0$
below 15~K, as obtained from a SQUID measurement. However, both
sets of experimental observations are reproduced to reasonable
accuracy upon introducing a model based on a broad distribution
of values of the nearest-neighbor coupling.

Complementing the
SQUID and pulsed fields measurements we have also performed
detailed $^{1}$H-NMR and $^{51}$V-NMR measurements. We find that the
temperature dependence of $\chi_0$  as estimated from the proton NMR
spectrum is in satisfactory agreement with that obtained from
our SQUID measurements.  From the $^{51}$V-NMR spectra measurements, a
high degree of inhomogeneity of the local environment of V ions
is suggested by the observation of a wide distribution of
quadrupole frequencies. Inhomogeneity of the local environment
is also suggested by the temperature dependence of the observed
line width and NMR shift of $^{51}$V-NMR. $T_1$ measurements
of both $^{1}$H and $^{51}$V reveal the existence of slow spin dynamics at
low temperatures. In particular, the fluctuation frequency of
the interacting system of V$^{4+}$ spins is found to show a power law
behavior, of the form $T^{3.5}$ at temperatures below 30~K, i.e., the
same behavior that has been found for many
antiferromagnetic rings and clusters with spins
$s>1/2$.\cite{BFL:06,RLB:PRB09}

\section*{Acknowledgment}

  This work was supported by the Deutsche Forschungsgemeinschaft
  through Research Unit 945.  The work at the Ames Laboratory
  was supported by the U.S. Department of Energy-Basic Energy
  Sciences under Contract No. DE-AC02-07CH11358. H.~N.  acknowledges the
  support by Grant in Aid for Scientific Research on Priority
  Areas (No.  13130204) from MEXT, Japan and by Shimazu Science
  Foundation. Computing time at the Leibniz Computing Center in
  Garching is gratefully acknowledged.
  We thank S. Bud'ko and P. C. Canfield for allowing us to use
  their MPMS system. 



\end{document}